\documentclass[pre,preprint,showpacs]{revtex4}

\usepackage[dvips]{epsfig} 
\usepackage{amsmath} 
\usepackage{amssymb}
\usepackage{amsfonts}
\usepackage{bm}

\begin{document}

\title{Molecular view of the Rayleigh-Taylor instability in compressible Brownian fluids}

\author{A. Wysocki}
\email{adam@thphy.uni-duesseldorf.de}

\author{H. L{\"o}wen} 
\email{hlowen@thphy.uni-duesseldorf.de}

\affiliation{Institut f{\"u}r Theoretische Physik II,
             Heinrich-Heine-Universit\"at D{\"u}sseldorf,
             Universit\"atsstra{\ss}e 1,
             D-40225 D\"usseldorf, 
             Germany}

\date{\today, in preparation for submission to {\sl Phys. Rev. E }}

\begin{abstract}

The onset of the Rayleigh-Taylor instability is studied  a compressible Brownian Yukawa fluid mixture on
the ``molecular'' length and time scales of the individual particles.
As a model, a two-dimensional phase-separated symmetric binary mixture of colloidal particles
of type $A$ and $B$  with a fluid-fluid interface separating an $A$-rich phase
from a $B$-rich phase is investigated by Brownian computer simulations
when brought into non-equilibrium via a constant external driving field
which acts differently on the different particles and perpendicular
to the interface. Two different scenarios
are observed which occur either for high or for low
interfacial free energies as compared to the driving force. In the first
scenario for high interfacial tension, the critical wavelength $\lambda_c$
of the unstable interface modes is in good agreement with the classical Rayleigh-Taylor
formula provided dynamically rescaled values for the interfacial tension
are used. The wavelength $\lambda_{c}$ increases with time representing
a self-healing effect of the interface due to a local density increase near the interface.
The Rayleigh-Taylor formula is confirmed even if $\lambda_c$ is of the order of a molecular correlation length.
In the second scenario for very large driving forces as compared
to the interfacial line tensions, on the other hand,
the particle penetrate easily the interface by the driving field and form
microscopic lanes with a width different from the predictions of the classical Rayleigh-Taylor formula.
The results are of relevance for phase-separating colloidal mixtures
in a gravitational or electric field.
    
\end{abstract}
\pacs{05.70.Ln, 05.70.Np, 82.70.Dd, 05.40.Jc}

\maketitle

\section{Introduction}

Interface instabilities in driven non-equilibrium systems are well-known
from macroscopic hydrodynamics \cite{Lehrbuch1,Lehrbuch2}. Examples include 
the classic Rayleigh-Taylor
instability \cite{Lehrbuch1} of a heavy liquid on top of another  lighter liquid,
the Mullins-Sekerka fingering instability \cite{Mullins1,Mullins2} 
in diffusive systems
and  the Saffmann-Taylor instability \cite{Saff1,Saff2} 
for compressed liquids of different viscosities.
However, what is much less clear is the  microscopic origin of these
instabilities, i.e.\ a view which resolves the microscopic discrete
particle trajectories causing the instability. Recently there
has been progress in simulating large systems with
10000-10000000 of discrete particles forming an interface
\cite{simulations1,simulations2,simulations3,Alder}. 
In particular, the microscopic origin of the
Rayleigh-Taylor instability was explored in huge molecular dynamics \cite{simulations1,simulations3,Alder}
and direct numerical \cite{simulations2} computer simulations. 
A natural question concerns the applicability of coarse-grained
hydrodynamics towards microscopic spatial dimensions.  
The basic quantities entering in
hydrodynamics auch as the viscosity or the surface tension 
are quantities which are only defined for large systems and exhibit
finite-size corrections when applied to small inhomogeneities.

Another approach to interfacial instabilities is via mesoscopic colloidal 
suspensions which bear the fascinating possibility to study
the particle trajectories  in real-space
by video microscopy in quasi-two-dimensional suspensions
\cite{Murray}. Well-characterized colloidal
suspensions serve as model systems for many questions of many-body systems
\cite{Pusey,myreview} including interfaces 
\cite{Lekkerkerker,Evans0,Velikov,Aarts,Hoogenboom1,Hoogenboom2}. 
The strength of an external field is easily tunable for colloids in striking
 contrast to molecular liquids.
The dynamics of the mesoscopic colloidal particles which are embedded in a microscopic solvent
is Brownian rather than Newtonian \cite{PRA1991} such that dynamical
quantities are different. To the best of our knowledge,
for compressible Brownian fluids nothing  is known about the onset of the Rayleigh-Taylor
instability on a length scale of interparticle distances. Clearly,
it will be different from molecular dynamics where heat will be generated
and inertia effects can lead to turbulence \cite{simulations1,simulations3,Alder}.
One promising investigation in the Newtonian case is for a suspension of 
heavy granular grains where Rayleigh-Taylor instabilities have recently 
been observed by Rehberg and coworkers \cite{Rehberg,Voeltz} but also in 
this work the individual particle trajectories were not resolved. 
Mixtures of colloids and polymers
with a real-space analysis of the colloids trajectories represent another
valuable system to look at for interfacial instabilities in external
driving fields \cite{Lekkerkerker}.

In the present paper we study, by Brownian dynamics computer simulations,
the particle-resolved onset of the Rayleigh-Taylor instability.
In our model, an interface separating an $A$-rich 
from an $A$-poor fluid is exposed to  an external field which 
 acts differently on the different particles and is directed  perpendicular
to the interface.
In order to keep the model simple \cite{Reichhardt} and to link to two-dimensional colloidal
suspensions, we  take two spatial
dimensions and a symmetric equimolar mixture interacting via
a Yukawa pair potential. Most of our characteristics, however, will
carry over to three-dimensional systems, to asymmetric mixtures, and
to different interparticle interactions.

The external field is so large that it
will induce local density inhomogeneities such that the fluid
is compressible in contrast  to the ordinary case treated within the
Navier-Stokes equations \cite{Alder}. 
We observe two different scenarios
which occur either for high or for low
interfacial free energies as compared to a typical thermal energy 
per average particle distance. In the first
scenario for high interfacial tension, an interfacial instability
is observed which is driven by the competition of the
cost in interfacial tension versus gain in potential energy,
similar in spirit to the classical Rayleigh-Taylor instability.
The classical threshold value for the wavelength of unstable
interface perturbations is confirmed provided the line
tension is dynamically rescaled in terms of the actual local density
at the interface.
The critical wavelength separating stable from unstable density
undulations increases with time. This can be interpreted as a
``self-healing effect'' of the interface caused by  a local density increase near the interface.
In the second scenario for small interfacial tensions, on the other hand,
the particle penetrate easily the interface by the driving field and form
microscopic lanes similar to previous simulation 
studies \cite{Netz,PRE02,JPCM02,Chakrabarti,Royal}. In this
case the classical threshold for the unstable wavelength is smaller
than a molecular correlation length such that  a breakdown of macroscale
hydrodynamics is expected.
Our  results are of relevance for phase-separating colloidal mixtures
in a gravitational or electric field where similar effects have recently been reported
by Aarts and coworkers \cite{Lekkerkerker} and for settling granular grains \cite{Rehberg}.

The paper is organized as follows: In section \ref{model}, we define the
model used and describe briefly our simulation technique. As a prerequisite,
a microscopic calculation of the density profile and the line tension is presented  in section \ref{equilibrium}.
We further review the classical Rayleigh-Taylor instability briefly in section \ref{classic}.
Results for the interface instability for high line tensions are presented in section \ref{hightension},
while the case of small line tensions is described in section \ref{lowtension}.
Conclusion are given in section \ref{conclusions}. In particular, we
discuss a possible verification of our predictions in experiments.

\section{The model}
\label{model}

In our model \cite{PRE02}, we consider a symmetric binary colloidal mixture comprising 
$N = N_{A}+N_{B}$ Brownian colloidal particles in $d = 2$ spatial dimensions.
The particles are in an area $S$  with a fixed total number density of $\rho = \frac{N}{S}$. 
Half of them are particles of type $A$, the other half is of type $B$ with  partial number density $\rho_{A} = \rho_{B} = \frac{\rho}{2}$.
The colloidal suspension  is held at fixed temperature $T$ via the  bath of microscopic solvent particles.
The colloidal particles of species $a$ and $b$ ($a,b\in\{A, B\}$)
are interacting via an effective Yukawa pair potential
\begin{equation}
\frac{V_{ab}(r)}{k_{B}T} = U_0\,\sigma_{ab}\,\frac{\exp\left[ -\kappa (r-\sigma_{ab})/\sigma \right]}{r},
\label{interaction}
\end{equation}
Here $r$ is the center-to-center separation, $k_{B}T$ is the thermal energy, $U_0$ is a dimensionless
amplitude, $\sigma$ is the particle diameter as a length scale and $\kappa$ is the inverse screening length.
The set of diameters $\sigma_{ab}$ is taken as 
\begin{equation}
\sigma_{AA}=\sigma_{BB}=\sigma
\label{diameter1}
\end{equation}
\begin{equation}
\sigma_{AB}=\sigma(1+\Delta)
\label{diameter2}
\end{equation}
We chose $\Delta>0$ corresponding to positive non-additivity. This implies that
the cross-interaction $V_{AB}(r)$ is  more repulsive than $V_{AA}(r)=V_{BB}(r)$, 
which drives phase separation into an $A$-rich and a $B$-rich phase.

The dynamics of the colloidal particles is overdamped Brownian motion. 
The friction constant $\xi = 3\pi \eta \sigma$ (with $\eta$ denoting the shear viscosity of the solvent) 
is assumed to be the same for both $A$ and $B$ particles. 
The constant external force for the $i$th particle of species $a$, $\vec F_{i}^{(a)}$, is acting differently on the both constituents of the binary mixture. 
It is $\vec F_{i}^{(A)}={\vec e_{z}}F$ for $A$ particles and $\vec F_{i}^{(B)}=-{\vec e_{z}}F$ for $B$ particles. 
The stochastic Langevin equations for the colloidal trajectories 
${\vec r}^{(a)}_{i}(t)$ with $i=1,...,N_a$ ($a\in\{A,B\}$) read as
\begin{equation}
\xi \frac{{d{\vec r}^{(a)}_{i}}}{dt} =
\sum_{\substack{j = 1\\j\not= i}}^{N_{a}} {\vec F}^{(aa)}_{ij}
+ \sum_{j = 1}^{N_{b}} {\vec F}^{(ab)}_{ij}
+ {\vec F}_i^{(a)}  
+ {\vec F}_i^{(\rm R)}(t). 
\label{langevin}
\end{equation}
where 
\begin{align}
{\vec F}^{(ab)}_{ij}=&-{\vec \nabla}_{{\vec r}^{(a)}_{i}} V_{ab}(r^{(ab)}_{ij}),
\label{force}
\end{align}
$r^{(ab)}_{ij}=\mid  {\vec r}^{(a)}_{i} - {\vec r}^{(b)}_{j}\mid$, 
and $b$ is the complementary index to $a$ ($b=A$ if $a=B$ and $b=B$ if $a=A$).
The right-hand-side includes all forces acting onto
the colloidal particles, namely the force resulting from
inter-particle interactions, the external constant force,
and the random forces ${\vec F}_i^{(\rm R)}$ describing 
the collisions of the solvent molecules with the $i$th 
colloidal particle. 
The latter are Gaussian random numbers with zero mean, $\overline {{\vec F}_i^{(\rm R)}}=0$,
and variance 
\begin{equation}
\overline{({\vec F}_i^{(\rm R)})_{\alpha }(t)({\vec F}_j^{(\rm R)})_{\beta }(t')}={{2k_BT}{\xi}}
\delta_{\alpha\beta} \delta_{ij}\delta(t-t').
\label{variance}
\end{equation}
The subscripts $\alpha$ and $\beta$ stand for the two 
Cartesian components.
In the absence of an external field and for $\Delta =0$, the model reduces to a two-dimensional  
Brownian Yukawa fluid in equilibrium which has 
been extensively investigated as far as structural and 
dynamical equilibrium correlations and freezing transitions
are concerned 
\cite{Loewen92,Loehle,Naidoo}. For a positive $\Delta$ and vanishing external drive,
our system will lead to equilibrium fluid-fluid phase separation including a critical point which has been
studied in non-additive hard-core models by theory and simulation, 
see e.g.\ Ref.\ \cite{Hoheisel,Nielaba,Lomba,Hamad,Giaquinta,Gozdz}.

We solve the Langevin equations of motion by Brownian dynamics simulations (BD) 
\cite{Hoffmann1,Hoffmann2,Roux} using a finite time-step $\Delta t$ and the technique of Ermak 
\cite{Allen,Ermak}. The typical size of the time-step was $0.003 \tau_{\rm B}$, where
$\tau_{B}=\xi\sigma^{2}/V_{0}$ is a suitable Brownian timescale.
As a reference, the field-free case ${\vec F}=0$ is studied extensively first.
This is an equilibrium situation. In the simulation set-up here, 
we put
$N_{A}=1000$ $A$ and $N_{B}=1000$ $B$ particles into a rectangular cell of 
lengths $L$ in $x$-direction  and $D$ in
$z$-direction  with
$\frac{D}{L}=\frac{8}{5}$ such that
the total colloidal number density is $\rho = \frac{N}{DL}$. Then the  system 
spontaneously exhibits a fluid-fluid interface along the $x$-direction separating an 
$A$-rich from a $B$-rich fluid provided the density is larger than the critical density.
To have an single interface system, without disturbing effects due to
external walls, periodic boundary conditions are used in the $x$-direction
while antiperiodic boundary conditions are used in $z$-direction. 
With antiperiodic boundary conditions, the particle type is changed
from $A$ to $B$ if an $A$ particle is crossing a boundary in
$z$-direction and vice versa. In a finite slab around this boundary,
all interactions are set to be equal, i.e. the cross interaction 
is $V_{AB}=V_{AA}=V_{BB}$. This avoids a second  interface
and minimizes finite-size effects.
Similar boundary condition shave been employed to study interfaces
of symmetric polymer mixtures by M\"uller, Binder and coworkers \cite{Muellerbox}.

We equillibrated $3\times 10^5$ time steps
 corresponds to an equilibration time of $900\tau_{B}$ and gathered statistics
again for a further simulation time of $900\tau_{B}$. We also checked
that our data do not suffer from finite size effects by choosing a larger system
with $N_{A}=4000$ $A$ and $N_{B}=4000$ $B$ particles with
$\frac{D}{L}=\frac{3}{5}$. The antiperiodic boundary conditions force the interface
to be parallel to the $x$-axis. All our investigations in nonequilibrium are
with this larger system size.

With an equilibrated interface as a starting configuration 
we suddenly turn on the external field which drives the particles
against the interface. An interface instability was observed. 
300 different equilibrated starting configuration were then typically used
in order to perform time-dependent averages. For our nonequilibrium simulations,
the antiperiodic boundary conditions in $z$ direction has no consequences once the particles
are driven against the interface.

\section{Density profiles and equilibrium fluid-fluid interfacial free energy}
\label{equilibrium}

In equilibrium, a phase-separated binary fluid mixture is characterized by its
$z$-dependent partial  density profiles $\rho_A (z)$ and $\rho_B (z)$ 
as defined via
\begin{equation}
\rho_a (z) = \left\langle\sum_{i=1}^{N_a} \delta (z-z^{(a)}_{i})\right\rangle
\label{density}
\end{equation}
where $a  \in\{A, B\}$ and $\langle\ldots\rangle$ denotes a canonical average.
Furthermore the associated
fluid-fluid interfacial free energy $\gamma$ is a key quantity. Since our model is two-dimensional, this
interfacial free energy $\gamma$ corresponds to a line tension. In a symmetric
binary mixture, the partial  density profiles and the
line tension $\gamma$ depend on the temperature and the total
number density $\rho$. 

Obviously, due to the $A$-$B$ symmetry of our model,
the density profiles are symmetric, i.e.\ $\rho_A(z) = \rho_B(-z)$, if the
interface position is at $z=0$. Computer simulation results 
of the density profiles for a strong
positive nonadditivity $\Delta =1.6$ and various bulk densities at a fixed
temperature are presented in Figure 1. For high densities one clearly 
idenfifies a depletion zone in the interface as
generated by the large non-additivity. For densities closer to the critical one, however,
the density profiles are getting flatter and particles interpenetrate mutually. 
For large bulk densities, a slight density
oscillation shows up as typical for fluid interfaces when the Fisher-Widom line
is exceeded \cite{Evans}. Clearly,
for the parameters chosen, the mixture
is almost completely phase-separated, i.e.\ the partial densities of the A 
particles is practically zero in the B-region and vice-versa. Again this
is different very close to the critical point.

We have also calculated the equilibrium
line tension as a function of density for fixed temperature by using  ``exact" computer
simulation techniques. The interfacial free energy can be gained by integration
of the anisotropy of the pressure tensor \cite{Buff,Smit,Chapela,Binder,Binder0}. 
In two spatial dimensions, this expression reads
\begin{equation}
\gamma=\int\limits_{-\infty}^{\infty} \left[P_{N}(z)-P_{T}(z)\right]\,dz
\label{linetension}
\end{equation}
where $P_{N}(z)=p_{zz}(z)$ and $P_{T}(z)=p_{xx}(z)$ are the diagonal components of the local pressure tensor.
This local pressure tensor has the following form \cite{Irving,Ono,Schofield,Rowlinson,Henderson}
\begin{equation}
\begin{split}
p_{\alpha\beta}(z)&= (\rho_{A}(z)+\rho _{B}(z))k_{B}T\delta_{\alpha\beta}\\
                   -\frac{1}{L}\Biggl\langle&\sum_{\substack{1\le i<j\le N_{a}\\a=A,B}}\left(\vec r^{(aa)}_{ij}\right)_{\alpha}\left({\vec F}^{(aa)}_{ij}\right)_{\beta}\frac{1}{ \left|z^{(aa)}_{ij}\right|}
                   \Theta\left(\frac{z-z^{(a)}_{i}}{z^{(aa)}_{ij}}\right)
                   \Theta\left(\frac{z^{(a)}_{j}-z}{z^{(aa)}_{ij}}\right)\\
                   \quad+&\sum_{\substack{1\le i\le N_{A}\\1\le j\le N_{B}}}\left(\vec r^{(AB)}_{ij}\right)_{\alpha}
                   \left({\vec F}^{(AB)}_{ij}\right)_{\beta}\frac{1}{\left|z^{(AB)}_{ij}\right|}
                   \Theta\left(\frac{z-z^{(A)}_{i}}{z^{(AB)}_{ij}}\right)
                   \Theta\left(\frac{z^{(B)}_{j}-z}{z^{(AB)}_{ij}}\right)\Biggr\rangle
\label{pressuretensor}
\end{split}
\end{equation}

where the subscripts $\alpha$ and $\beta$ stand for the two Cartesian components, ${\vec r}^{(ab)}_{ij} = {\vec r}^{(a)}_{i} - {\vec r}^{(b)}_{j}$, 
$z^{(ab)}_{ij} = z^{(a)}_{i} - z^{(b)}_{j}$ and $\Theta(z)$ is the Heaviside step function.
The $z$-dependence of the anisotropy of the pressure tensor is plotted in Figure 1 as well.
As can be deduced from Figure 1, its main weight  is centered in the interface position around $z=0$.
The oscillations of the density field for high densities correlate to that of the pressure tensor.

Simulation results for the density-dependent line tension $\gamma ( \rho )$ for two different nonadditivities
$\Delta=0.8$ and $\Delta =1.6$ are presented in Figure 2.
Most of our calculations were for parameter combinations well away from the critical
point where the line tension is mainly governed by internal energy such that
the anisotropy of the pressure tensor is significant resulting in a 
relative small statistical error for $\gamma$, see Figure 2a.
Some further points are also for smaller densities closer to the critical point,
see Figure 2b.
By crudely extrapolating the data one could estimate the critical density to
be at $\rho_{c }\sigma^2\approx 0.08$ for $\Delta=0.8$ and 
$\rho_{c }\sigma^2\approx 0.04$ for $\Delta=1.6$ as indicated  by arrows in Figure 2b.
This extrapolation was 
cross-checked by mapping our Yukawa system
onto that of effective non-additive hard disks using the prescription 
of Barker-Henderson for the effective hard-core diameter \cite{HansenMcDonald}.
The critical point can then be read off from that 
of a non-additive symmetric hard disk binary mixture which has been  studied
in detail via theory and simulation by Giaquinta and coworkers \cite{Giaquinta}.
We find good agreement with our extrapolation as compared to the mapping procedure.
As can be further deduced from Figure 2, the line tension 
$\gamma ( \rho )$ is strongly increasing with density. This is expected as 
a density increase means a smaller spacing between A and B particles such that
the energetic non-additivity is getting more pronounced. 
At fixed density and temperature, the line tension increases with $\Delta$
which is clearly due to the fact that a larger nonadditivity leads to a larger
energy cost of different particle species meeting
at the interface.

\section{Classical Rayleigh-Taylor instability}
\label{classic}

The classical Rayleigh-Taylor instability is obtained for  a heavy incompressible
liquid on top of a lighter incompressible liquid \cite{Lehrbuch1,Sharp}. A small
harmonic interfacial undulation with a wave length $\lambda$ yields a favorable decrease
in potential energy but at the same time a free energy penalty due to the increasing arc length
of the interface which costs line tension. If the wave length is larger than a 
critical one $\lambda > \lambda_c$, however, the penalty is smaller  
than the gain such that an unstable
mode whose amplitude  grows  in time is present.
The critical wave length can be calculated in our two-dimensional situation as 
\begin{equation}
\lambda_{c} = \frac{2\pi}{k_c} = 2\,\pi\,\sqrt{\frac{\gamma}{\bigl|\vec F^{(A)}-\vec F^{(B)}\bigr|\rho}}
\label{rtlength}
\end{equation}
where $\gamma$ is the line tension as introduced in the previous section.
Since the concept of line tension is a macroscopic one,  Eq.(\ref{rtlength})
is only justified as long as $\lambda_c$ is much larger than any microscopic
distance as e.g. the mean interparticle spacing $a=\frac{1}{\sqrt{\rho}}$. This requires  that
the line tension has to be  much larger than the driving force difference 
\begin{equation}
\lambda_{c}\,\gg\,a\quad\Rightarrow\quad\gamma\,\gg\,\left|\vec F^{(A)} - \vec F^{(B)}\right|
\label{limit1}
\end{equation}
In the opposite limit
\begin{equation}
\gamma\,\ll\,\left|\vec F^{(A)} - \vec F^{(B)}\right|
\label{limit2}
\end{equation}
it is expected that the classical concept of the Rayleigh-Taylor instability will break down.
We shall explore this case in detail in section VI. Another question concerns the
applicability of the Rayleigh-Taylor instability criterion when $\lambda_c$ is of
a similar order than the mean interparticle spacing $a$. As we shall show
below, the criterion works remarkable well even if $\lambda_c$ is close to $a$.

\section{Results for high and moderate interfacial tensions}
\label{hightension}
Typical snapshots of our nonequilibrium computer simulations are presented in Figure 3
for 4 different times and for two different densities $\rho^*=\rho\sigma^{2}=0.2,0.4$
and two different strengths of the driving force $F^*=F\sigma/k_{B}T=10,40$.
The parameter combinations investigated and the
corresponding ratios of the line tension $\gamma$ and the driving force $F$ 
are summarized in Table I. 
They are of the order 1 (for combination A-C) or a bit smaller (for
combination D) such that 
 the Rayleigh-Taylor instability wavelength is larger than or of the order
of the mean interparticle spacing $a$, see again Table I for the ratio $\lambda_c/a$.
We call an interface tension ``high'' if $\lambda_{c}/a > 1$ and ``intermediate''
if $\lambda_{c}/a\approx 1$.
This is in contrast to the case of low surface tensions (combination E)
where $\lambda_c/a$ is significantly smaller than 1 which is studied in chapter VI.

The system was started with an equilibrated single interface situation. 
The initial interface is pretty smooth but carries
capillary fluctuations on it. The density depletion
shown in Figure 1, manifests itself as  a clear void region in the interface 
reminiscent of a ``forest-aisle'' which is induced by the strong energy cost
when two different particle species do meet.

Then instantaneously the external driving forces were turned on forcing the
particles to drift against each other. This first causes a local density increase close 
to the interface. Then interface undulations are getting more pronounced deforming the flat
interface. This is the most efficient channel for different particle species
 to reverse their height. A typical characteristic undulation wave length can be identified
from the simulation snapshot which we shall quantify later. Furthermore
nonlinear effects such as interfacial overhangs (``mushrooms'') can be seen 
(e.g.\ in Figure 3a for the largest time) until the mixture
penetrates through eachother and the height reversal is complete. 

In order to quantify the structural signatures of the interface instability further,
we consider the Fourier transform of the interface position as a function of time.
In detail, let $h(x, t)$ be the interface position of a given configuration at a time $t$, and
\begin{equation}
{\tilde h} (k,t)= \int\limits_{0}^{L} h(x, t)\exp{(-ikx)}\,dx
\label{Fourier}
\end{equation}
the  corresponding Fourier transform. 
We are interested in the time  dependence of the averaged  power
spectral density which is defined as 
\begin{equation}
P(k,t) = \langle{\tilde h} (k,t){\tilde h}^{*} (k,t)\rangle
\label{psd}
\end{equation}
where $\langle\ldots\rangle$ now denotes an ensamble average over inital equilibrated configurations
which dynamically evolved after a time $t$. If, at a given $t$, $P(k,t)$
possesses a sharp maximum at $k=k_m$, this implies that the interface will exhibit
mainly undulations with a wave length of $\lambda_{m} = 2\pi/k_{m}$.

As  an aside, let us remark that an alternative way of obtaining the line
tension $\gamma$ is via the
equilibrium capillary wave spectrum $P(k,t = 0)$ which behaves as 
\begin{equation}
P(k,t = 0) = \frac{k_{B}T}{\gamma}\frac{L}{k^2}
\label{capillar}
\end{equation}
for small wave vectors $k$ \cite{Lovett,Weeks,Rowlinson}. We have checked that values
for the line tension $\gamma$  as obtained from this
formula are consistent with those
obtained from Eq.(\ref{linetension}).
 
We are now in a position to define a differential growth rate $\Gamma (k,t)$ via
\begin{equation}
\Gamma (k,t) = \frac{1}{P(k,t)}\,\frac{d}{dt}P(k,t)
\label{growthrate}
\end{equation}
A positive sign of $\Gamma (k,t)$ implies that a mode of wave number $k$ is growing
at a time $t$ while a negative sign means that - at a given time $t$ - the mode is decreasing.

In detail, the following numerical procedure was used to obtain the actual interface position $h(x,t)$: 
For a given particle configuration at time $t$, we construct Voronoi cells around each particle and
define the interface based on the associated Voronoi polygons. 
The polygon vertices that belong to cells of particles of both species $A$ and $B$ define  a set of $M^{*}$
non-equidistant co-ordinates $\left(h_{n},x_{n}\right)_{n=1,\ldots,M^{*}}$.
Now we eliminate the protuberance in the interface profile to get an functional interrelation for $h(x)$.
For further analysis we interpolate $\left(h_{n},x_{n}\right)_{n=1,\ldots,M^{*}}$ by a cubic spline and 
eliminate sharp bends by using a simple low-pass filter
\begin{equation}
h_{n}\rightarrow h_{n}^{*} = \left(h_{n}+h_{n+1}\right)/2
\label{l-p-filter}
\end{equation}     
where $n=1, \ldots,M$ and $h_{M+1}\equiv h_{1}$. This procedure results in a unique $z$-position $h(x,t)$ of the interface
at a given time $t$.

Results for the differential growth rate $\Gamma (k,t)$ are shown in Figure 4 as a contour 
plot in the plane spanned
by the time $t$ and the wave vector $k$. As in Figure 3, data for the four
different parameters combinations A-D are presented. 
The white {\bf zero line} summarizes
points where $\Gamma (k,t)$ vanishes and separates two regimes with growing (stable)
modes and damped (unstable) wave numbers. Although this line
is noisy, it can be read off from Figure 4 that  there is a whole band of wave numbers which
are growing after an induction time of roughly $\tau_B$. 
The lower limit of unstable wave numbers is significantly larger than the inverse simulation box length,
while its upper limit is always smaller than a microscopic wave number of the order of $2\pi/a$.
Between these two boundaries, roughly at the arithmetric mean of the two limits, there is a 
wave number with a {\bf maximal} growth rate.
!!
The first basic observation is that for parameter combinations A-C where $\lambda_c/a$ is larger than 1 the upper limit of unstable wave numbers is {\it decreasing} with time.
!!
This has to do with our equilibrated starting configuration. When the external field is turned on,
there is a sedimentation-like process towards the interface which yields a local
density increase at the interface. This behaviour can directly be seen in the simulation snapshots of Figure 3.
Thereby the interface is getting stiffer as a fuction of time. We call this important effect
{\it self-healing} of the interface, i.e. as a function of time the interface is getting less vulnerable
with respect to short wavelength undulations.  A 
possible destruction of the interface is efficiently blocked by a density accumulation.
It is tempting to correlate this to
an effective interface tension with a scaled density as obtained e.g.\ by the maximal
total density $\rho_m$ close to the interface. This maximal density $\rho_m$ is a function of time.
If one plugs this time-dependent density into the expression for $\gamma (\rho )$, one
obtains a dynamically rescaled Rayleigh-Taylor expression (9).
The corresponding wave number is also shown in Figure 4, see the open  circles with error bars.
Clearly the basic effect is encaptured by the dynamical rescaling as the
classical Rayleigh-Taylor unstable wave number coincides well with the upper unstable limit.
The agreement  is in particular encouraging for parameter combinations A-C where $\lambda_c/a > 1$.

A special remark is in order for parameter combination D where $\lambda_c$ is of the order of $a$.
In fact, we estimate a  microscopic wave number  by $2\pi\sqrt{\rho_{m}}$ and
data for this wave vector $2\pi\sqrt{\rho_{m}}$ are included in Figure 4 as crosses. 
Even in this case, the comparison between the upper unstable wave number and
the scaled  Rayleigh-Taylor prediction  is qualitative. This implies that the 
classic Rayleigh-Taylor criterion is astonishingly  robust even close to molecular spacings.

On the other hand, the {\bf maximal} growth rate is seen as light region for small wave vectors.
This wave number is slightly decreasing with time as well. We have further compared
the wave number of maximal growth by calculating the second maximum of the height-height-correlation function which
is defined as 
\begin{equation}
C(k, t) = \left\langle h(x, t) h(x+\frac{2\pi}{k},t)\right\rangle
\label{acf}
\end{equation}
The position of the second maximum of $C(k,t)$ is also included as a square-line in Figure 4.
It is increasing as a function of time towards larger wave numbers, i.e.\ smaller
wave lengths. The increase is small and confined to a small ``induction time'' of the process.
We attribute this to an initial process which has to do with the fact that due to the
non-homogeneous capillary wave spectrum Eq.(\ref{capillar}), undulations with larger wave lengths have initially 
a larger amplitude and have  therefore more weight in $C(k,t)$. This gives rise to the crossover
to higher wave number after an induction time which is pretty sharp for the parameter combinations A and B.

Finally we note that at higher $k$ the growth rate is getting positive for larger times.
This is due to the fact that the interface position is no longer sinoidal but starts
to exhibit sharp parts close to overhangs. This is turn will generate higher Fourier modes
to grow and this is what is indicated by the zero-line at higher $k$ and larger times.

We finally think that the zero-line at high $k$ and smaller times (in Figure 4a and 4b) is statistical noise.
Furthermore,
a full comparison of the differential growth rates to a hydrodynamic
approach \cite{Lehrbuch1} requires a dynamically scaled viscosity
and is left for future studies.

\section{Results for low interfacial tensions}
\label{lowtension}

Similar computer simulations were done as in the previous section but now in the regime
where the line tension was low in the sense of the criterion of Eq.(\ref{limit2})
such that $\lambda_c$ was significantly smaller than $a$. Simulation
snapshots are shown for $\rho^*=0.075$ close to the estimated critical density in Figure 
5. In this case, the interface is penetrable by the strongly driven particles
and $\gamma/F=0.002$ and $\lambda_c/a=0.2$, see the parameter combination E in Table I.

In penetrating the interface, the particles form lanes similar to the behaviour in additive 
mixtures in Refs. \cite{PRE02,JPCM02}. The width
of the lanes is comparable to the correlation length of the mixture meaning
that single worms of particles are formed \cite{Chakrabarti}. At the head
of the lanes both spikes and extrusions are visible. A similar
behaviour is found for even smaller densities below the critical density where
the system is mixed in equilibrium.
Hence the Rayleigh-Taylor criterion would predict a submolecular unstable wave length
but the actual realized instability wavelength is a 
molecular correlation length of the system of the order of $a$.

\section{Conclusions}
\label{conclusions}

In conclusion, we have studied the onset of the Rayleigh-Taylor
 interface instability on length
scales of interparticle distances for a compressible Brownian fluid mixture. 
Depending on the equilibrium interfacial tension, either a direct transition towards lane
formation or the macroscopic Rayleigh-Taylor instability was found.
An interesting dynamical effect is the self-healing mechanism for the interface
which is produced by a density accumulation of particles driven against eachother which then
causes an increase of the instability wavelength as a function of time.

It would be interesting to study the dynamical process of phase separation by starting
from a completely mixed system under the influence of the driving forces. Here
one would expect a subtle interplay between phase separation kinetics which typically results
in fractals and lane formation \cite{Sendai}. One should further investigate more
extensively the dependence on the ``sedimentation height'', i.e.\ how the data are affected by a 
larger initial $D$ in $z$-direction perpendicular to the interface.
Moreover. it would be interesting to start the simulation by impressing a prescribed 
wave length as an interface undulation and checke its differential growth. A third
set-up for another initial configuration which is typically used
for the molecular dynmics simulations \cite{Alder} would be to equilibrate first with respect
to the external field with a fixed interface as induced e.g.\ by a thin hard platelet
 separating the two fluid phases. The effects of different starting configuration
on the onset of the Rayleigh-Taylor instability 
will be left for future studies.

For simplicity, all our simulations were done in two spatial dimensions.
In three dimensions, similar effects should persist. However, the characteristic wavelength
has now two components parallel to the interface. The self-stabilizing effect of the interface and
the two extreme limits should be similar as in two dimensions. The lane formation, for
instance, has also be shown to be present in three spatial dimensions.

Let us finally discuss some possible experimental verifications. Strongly non-additive mixtures
which phase-separate into two fluid phases are found in colloid/polymer mixtures which 
exhibit colloid-rich and colloid-poor fluid phases for size ratio of about 0.5 or 
larger between the polymer 
and the colloid  \cite{Dijkstra}. Recently a phase-separating mixture of colloids and polymers
was observed in sedimentation by Aarts and coworkers \cite{Aarts}. In particular, 
it was observed that large regions of phase-separated
colloid-rich and colloid-poor phases exhibit a transition towards lane formation. 
The characteristic
wavelength is of the order of that given by the Rayleigh-Taylor instability Eq.(\ref{rtlength})
\cite{Aarts3}.
A full quantitative comparison has still to be performed. Another system were such lane
formation has been seen and which could be a candidate for a quantitative comparison
is a  xanthan-colloid mixture \cite{Aarts2}. 
Finally we mention the fascinating possibility to study the mixture of complex plasmas
involving dust grains for interface instability in real space \cite{Morfill0,Morfill}.

\section*{Acknowledgments}
We  thank D. G. A. L. Aarts, B. J. Alder, K. Binder, R. Blaak, G. Morfill, K. H. Spatschek,
K. R. Mecke  and V. Steinberg for helpful remarks.
Financial support from the DFG (Sonderforschungsbereich TR6) is gratefully
acknowledged.

\newpage


\newpage

\begin{center}
{\large FIGURE AND TABLE CAPTIONS}
\end{center}

Figure 1: 
Averaged density profiles $\rho_A(z)$ and $\rho_B(z)$ of the two particle species 
and anisotropy $P_N-P_T$ of the pressure tensor
as a function of the distance $z$ perpendicular to the interface for $\Delta =1.6$ and for
four different densities: $\rho\sigma^2=0.4$ (solid lines), $\rho\sigma^2=0.3$ (dashed lines),
$\rho\sigma^2=0.2$ (dot-dashed line) and $\rho\sigma^2=0.15$ (short-dashed lines).

Figure 2: 
Reduced line tension $\gamma \sigma / k_BT$ as a function of reduced bulk density
$\rho \sigma^2$ for two different nonadditivities $\Delta =0.8$ (solid line) and 
$\Delta =1.6$ (dot-dashed line). (a) Away from the critical point. (b)
Closer to the critical point.
The estimate for the critical density are shown as arrows.

Figure 3:
Typical simulation snapshot for an interface in nonequilibrium with an external drive.
There is a sharp interface between an $A$-rich and a $B$-rich fluid phase. The starting
configuration was an equilibrated interface at $z=0$.
Four different times are shown as given in the legend.
a) $\rho\sigma^2=0.2$ and $F^*=F\sigma/k_{B}T=10$,
b) $\rho\sigma^2=0.2$ and $F^*=F\sigma/k_{B}T=40$,
c) $\rho\sigma^2=0.4$ and $F^*=F\sigma/k_{B}T=10$,
d) $\rho\sigma^2=0.4$ and $F^*=F\sigma/k_{B}T=40$.

Figure 4:
Contour plots of the growth rate $\Gamma (k,t)$. Zero growth is shown by the white line.
The rescaled classical Rayleigh-Taylor wave length $2\pi/\lambda_c$ is shown by the circles.
Here is time-dependent maximum $\rho_{m}$ of the total density at the interface is used.
The statistical error stems from the uncertaincy of this density. The position
of the second maximum in the height-height correlation function are the squares.
The corresponding microscopic wave number as defined by $2\pi  \rho_{m}^{1/2}$
and gives an estimate for the threshold to where a microscopic length scale are realized.

Figure 5:
Typical simulation snapshot for an interface in nonequilibrium with an external drive.
There is a sharp interface between an $A$-rich and a $B$-rich fluid phase. The starting
configuration was an equilibrated interface at $z=0$.
Four different times are shown as given in the legend. The parameters are
 $\rho\sigma^2=0.075$ and $F^*=F\sigma/k_{B}T=80$.

Table 1:
Summary of the 5 parameters combinations studied in the paper.
The corresponding ratios $\gamma/F$ and $\lambda_{c}/a$ are also given.

\newpage
\begin{tabular}{|p{1.2cm}|p{1.2cm}|p{1.2cm}|p{1.2cm}|p{1.2cm}|}\hline
\text{}&\textbf{$\rho^{*}$}&\textbf{$F^{*}$}&\textbf{$\gamma/F$}&\textbf{$\lambda_{c}/a$}\\ \hline\hline
A&0.4&10&1.3&5\\ \hline
B&0.4&40&0.32&2.5\\ \hline
C&0.2&10&0.2&2\\ \hline
D&0.2&40&0.05&1\\ \hline
E&0.075&80&0.002&0.2\\ \hline
\end{tabular}

\newpage
\begin{center}
\begin{figure}
\epsfig{figure=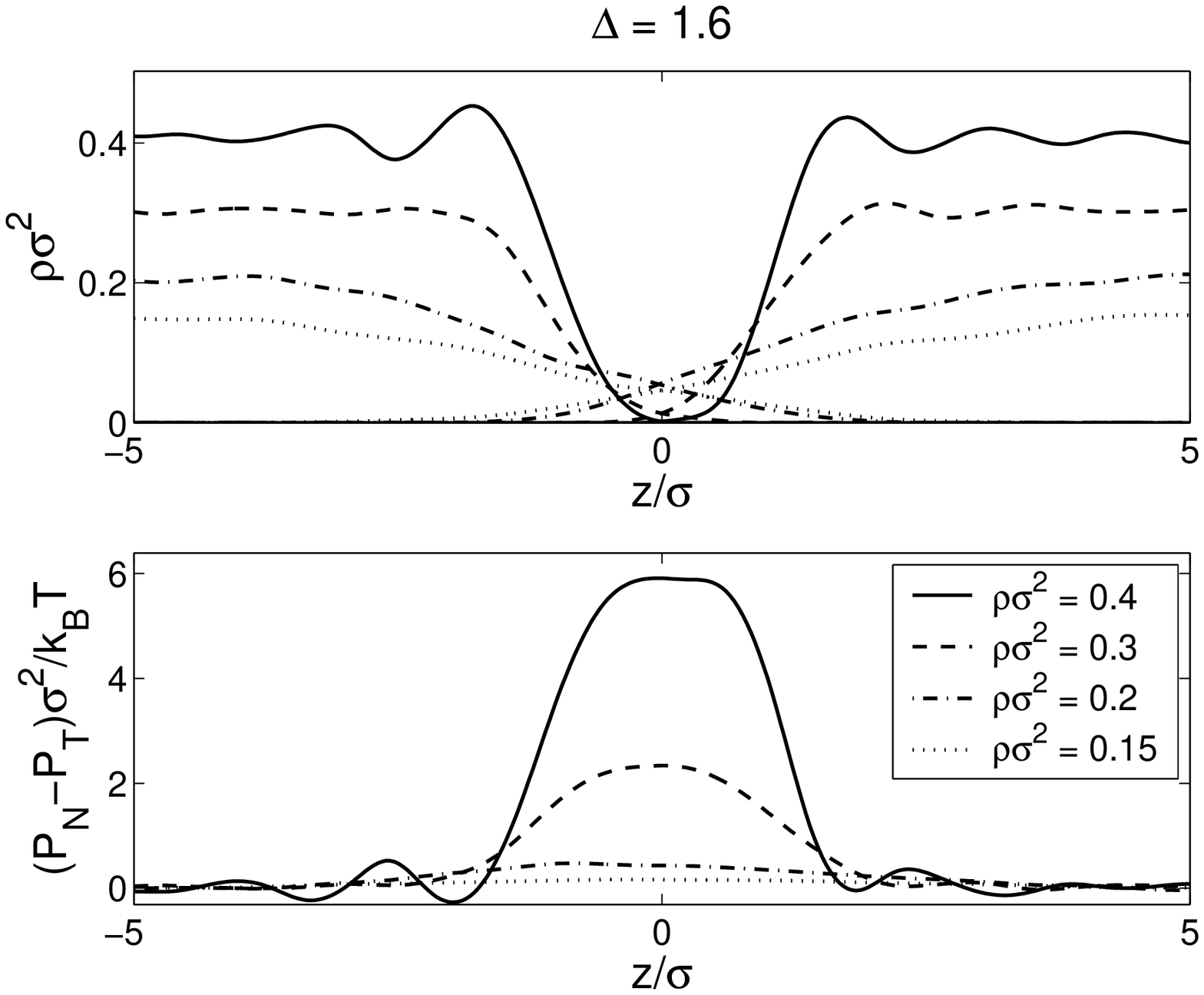,width=14cm}
\label{Figure 1} 
\end{figure}
\end{center}

\newpage
\begin{center}
\begin{figure}
\epsfig{figure=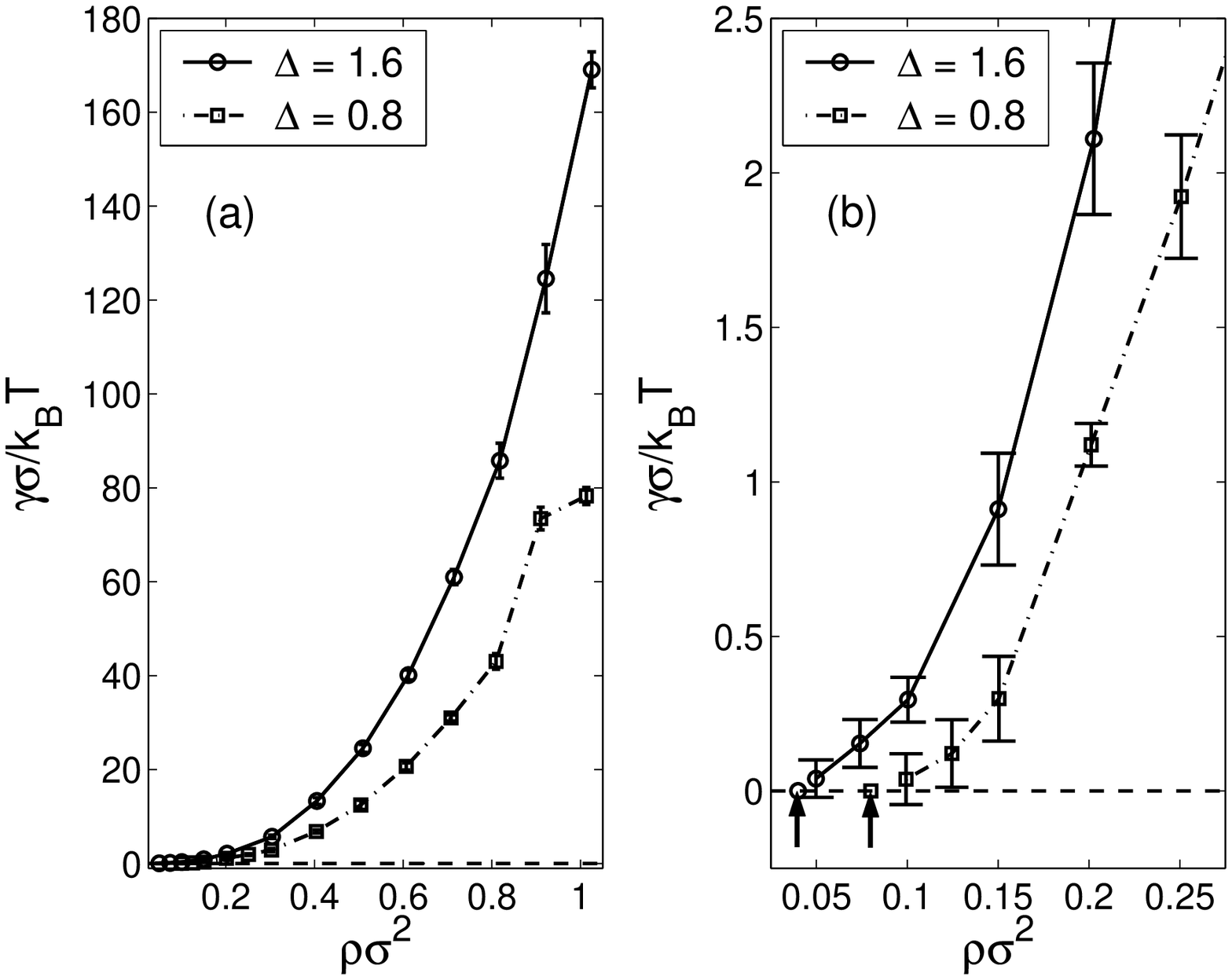,width=14cm}
\label{Figure 2} 
\end{figure}
\end{center}

\newpage
\begin{center}
\begin{figure}
\epsfig{figure=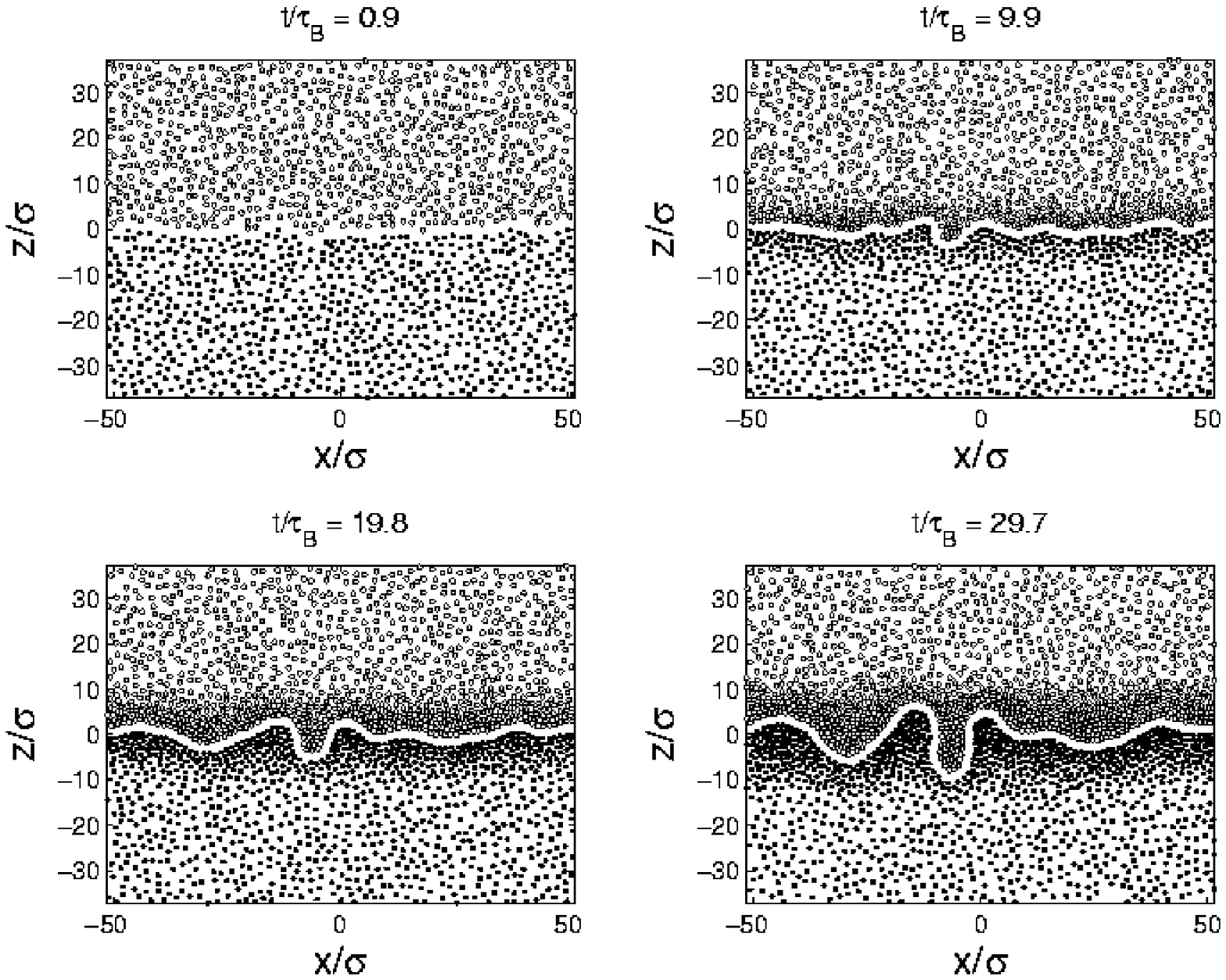,width=14cm}
\label{Figure 3a} 
\end{figure}
\end{center}

\newpage
\begin{center}
\begin{figure}
\epsfig{figure=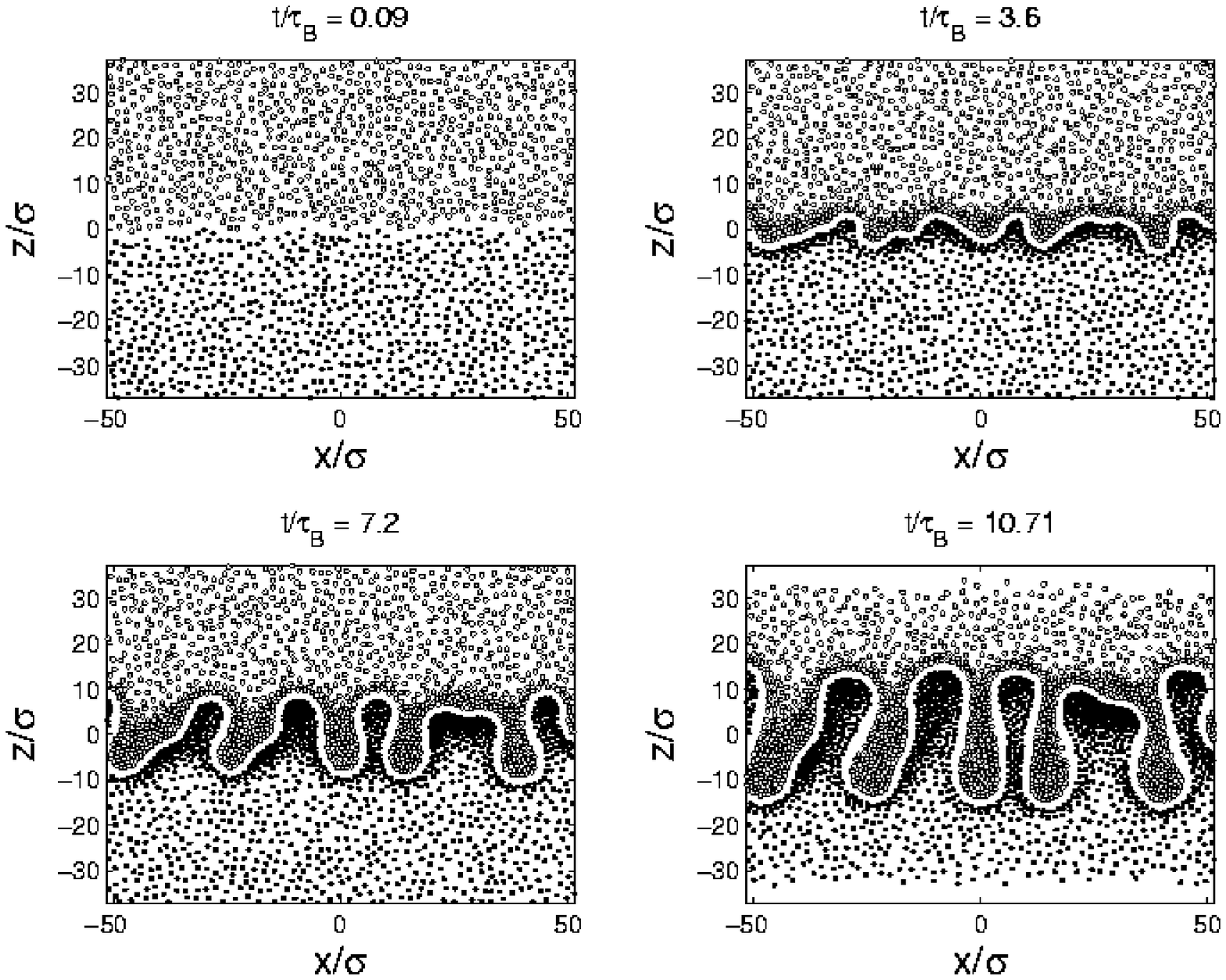,width=14cm}
\label{Figure 3b} 
\end{figure}
\end{center}

\newpage
\begin{center}
\begin{figure}
\epsfig{figure=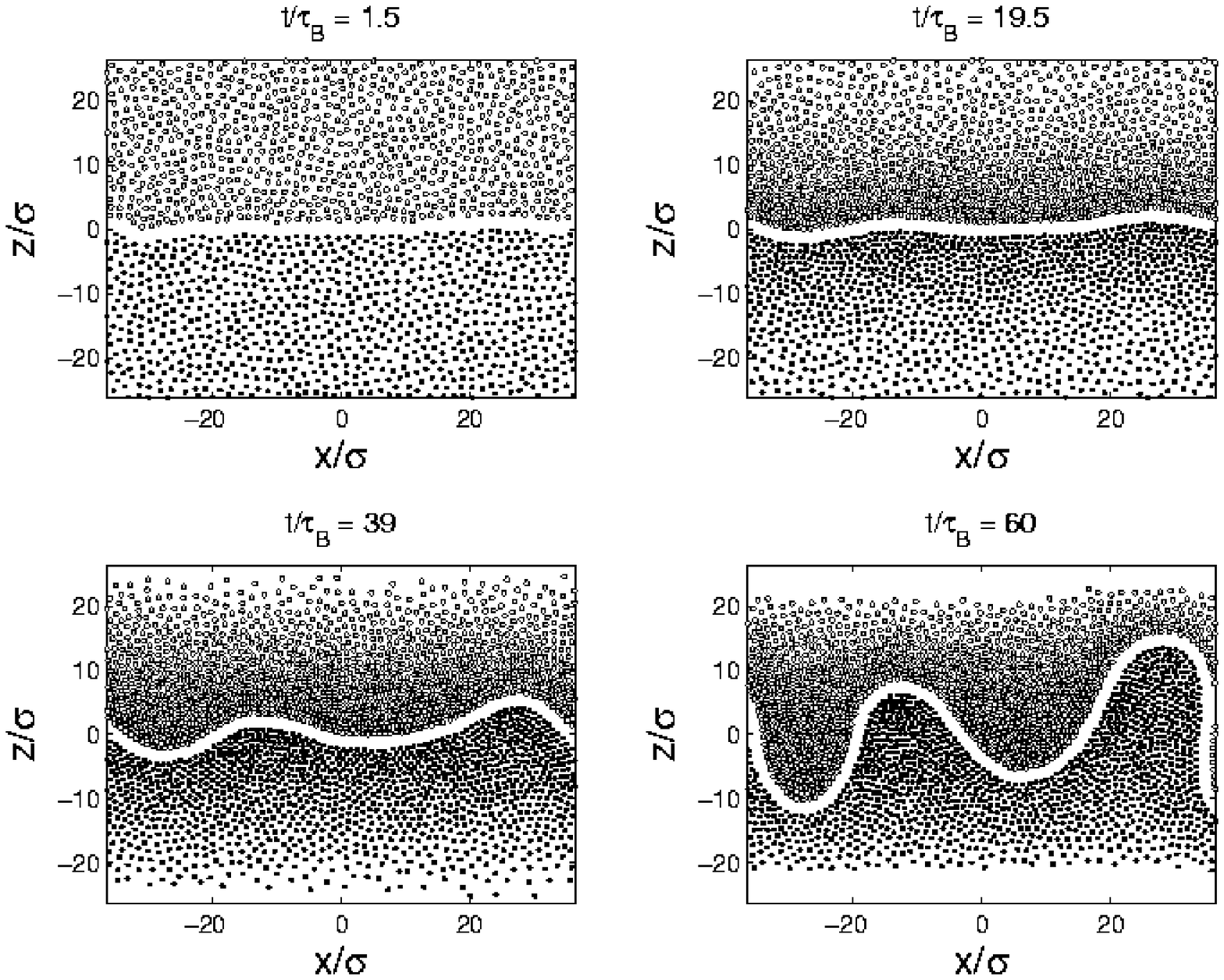,width=14cm}
\label{Figure 3c} 
\end{figure}
\end{center}

\newpage
\begin{center}
\begin{figure}
\epsfig{figure=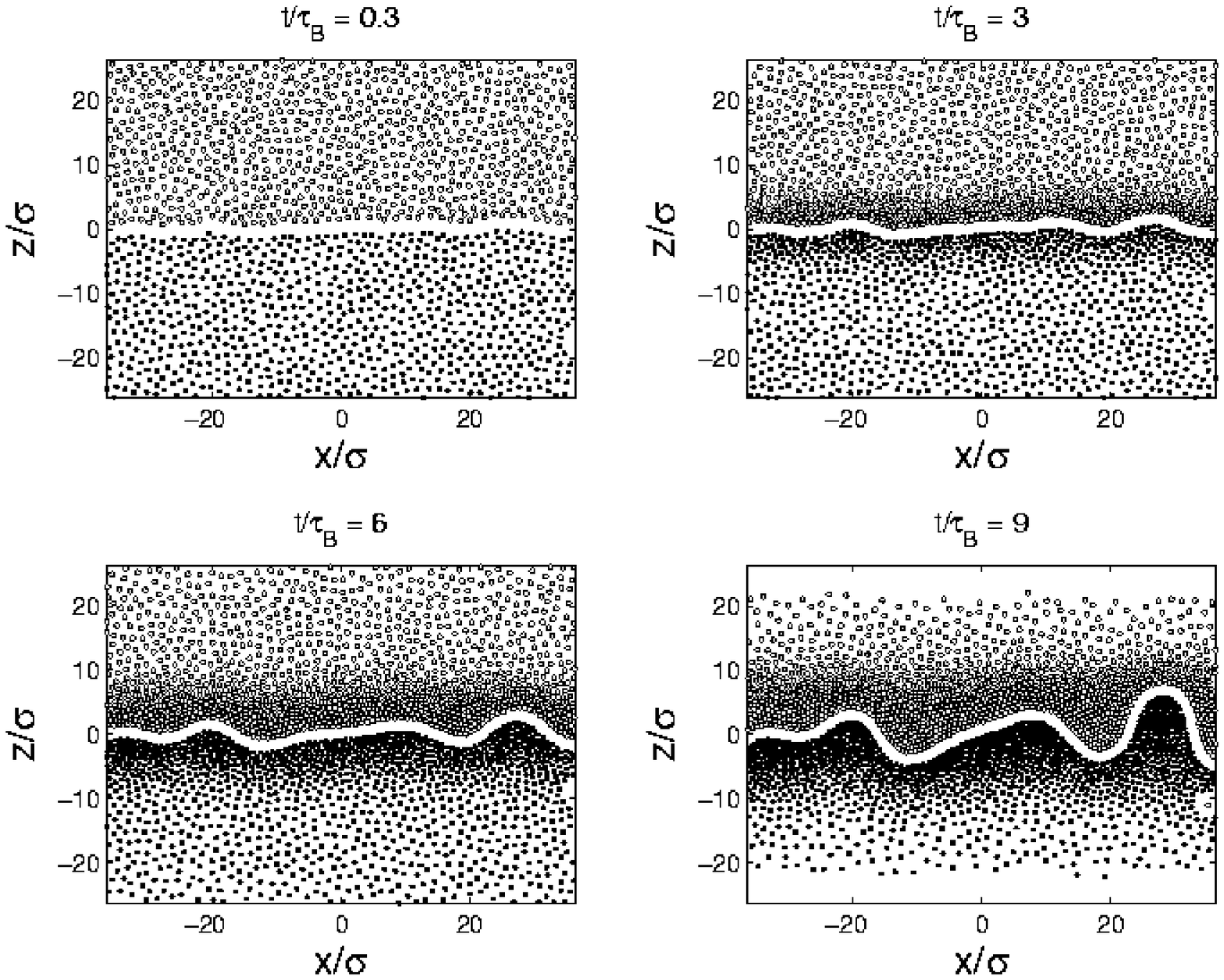,width=14cm}
\label{Figure 3d} 
\end{figure}
\end{center}

\newpage
\begin{center}
\begin{figure}
\epsfig{figure=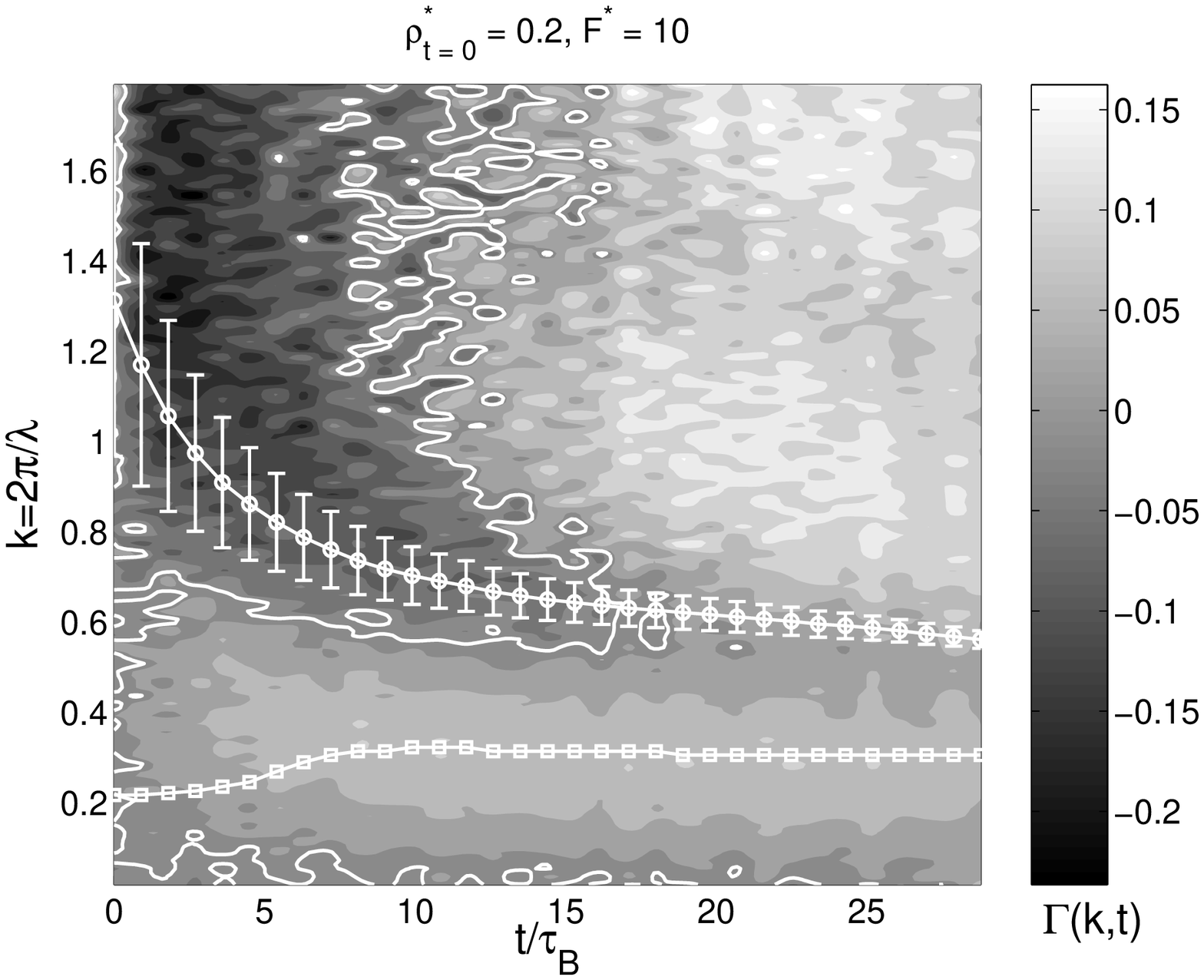,width=14cm}
\label{Figure 4a} 
\end{figure}
\end{center}

\newpage
\begin{center}
\begin{figure}
\epsfig{figure=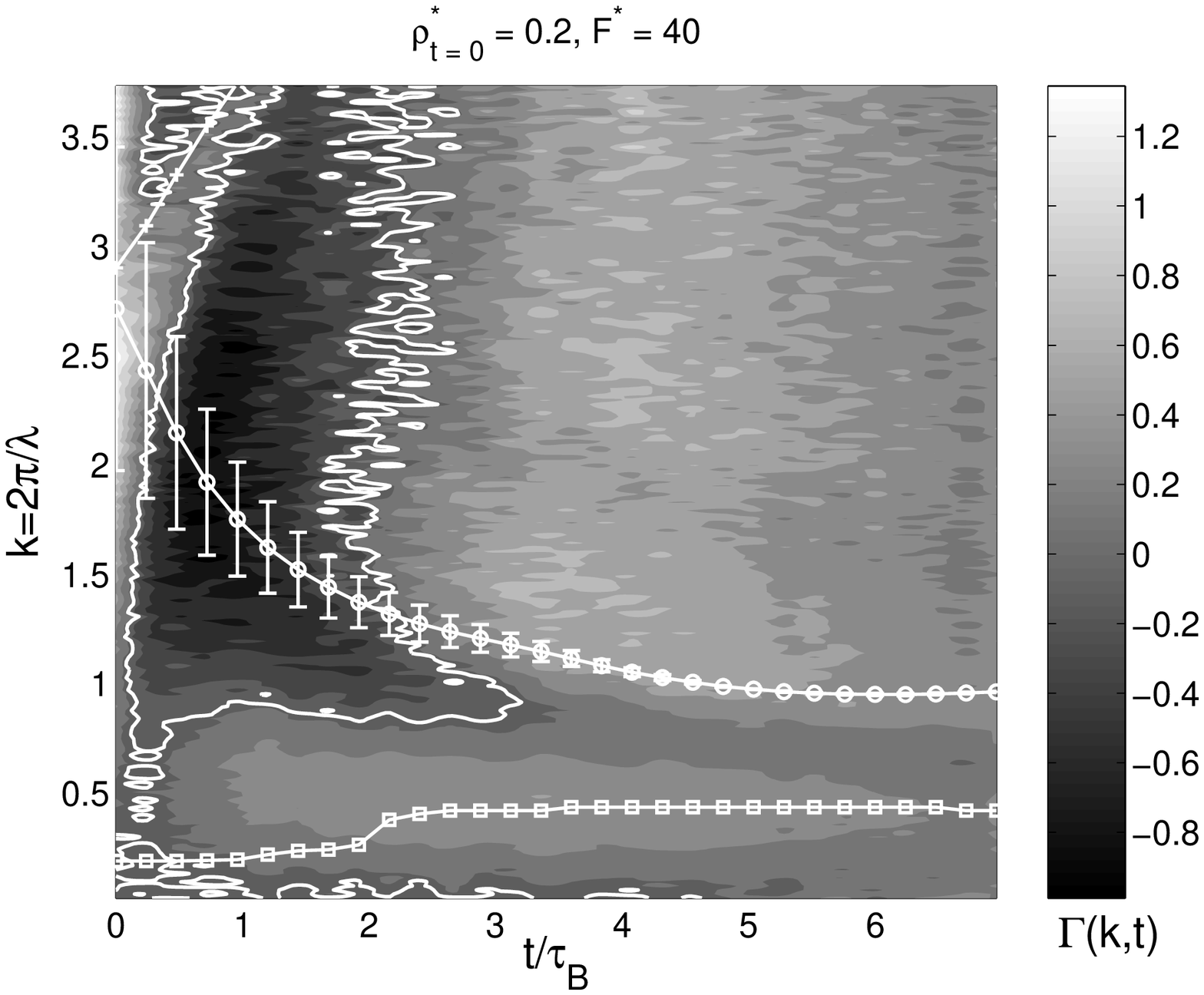,width=14cm}
\label{Figure 4b} 
\end{figure}
\end{center}

\newpage
\begin{center}
\begin{figure}
\epsfig{figure=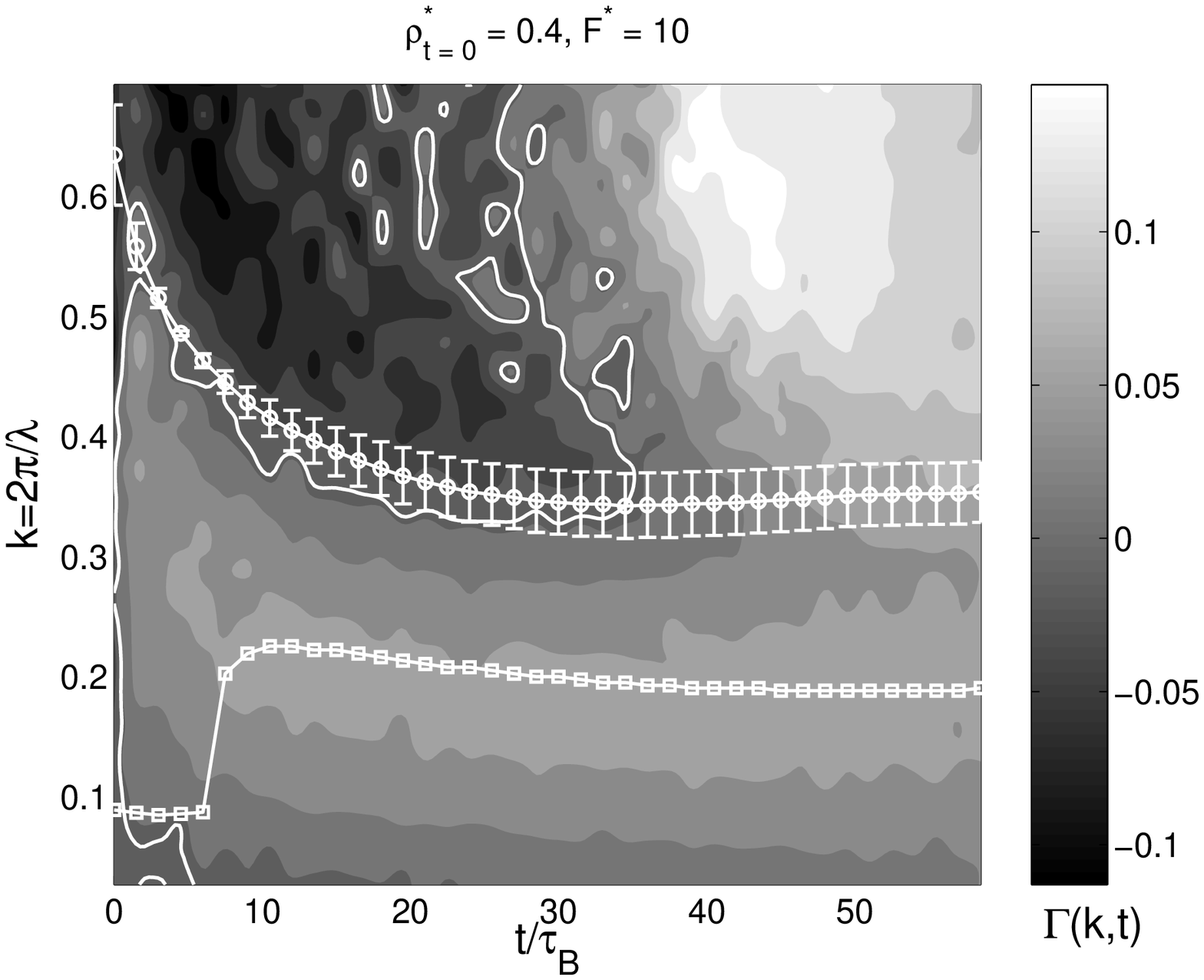,width=14cm}
\label{Figure 4c} 
\end{figure}
\end{center}

\newpage
\begin{center}
\begin{figure}
\epsfig{figure=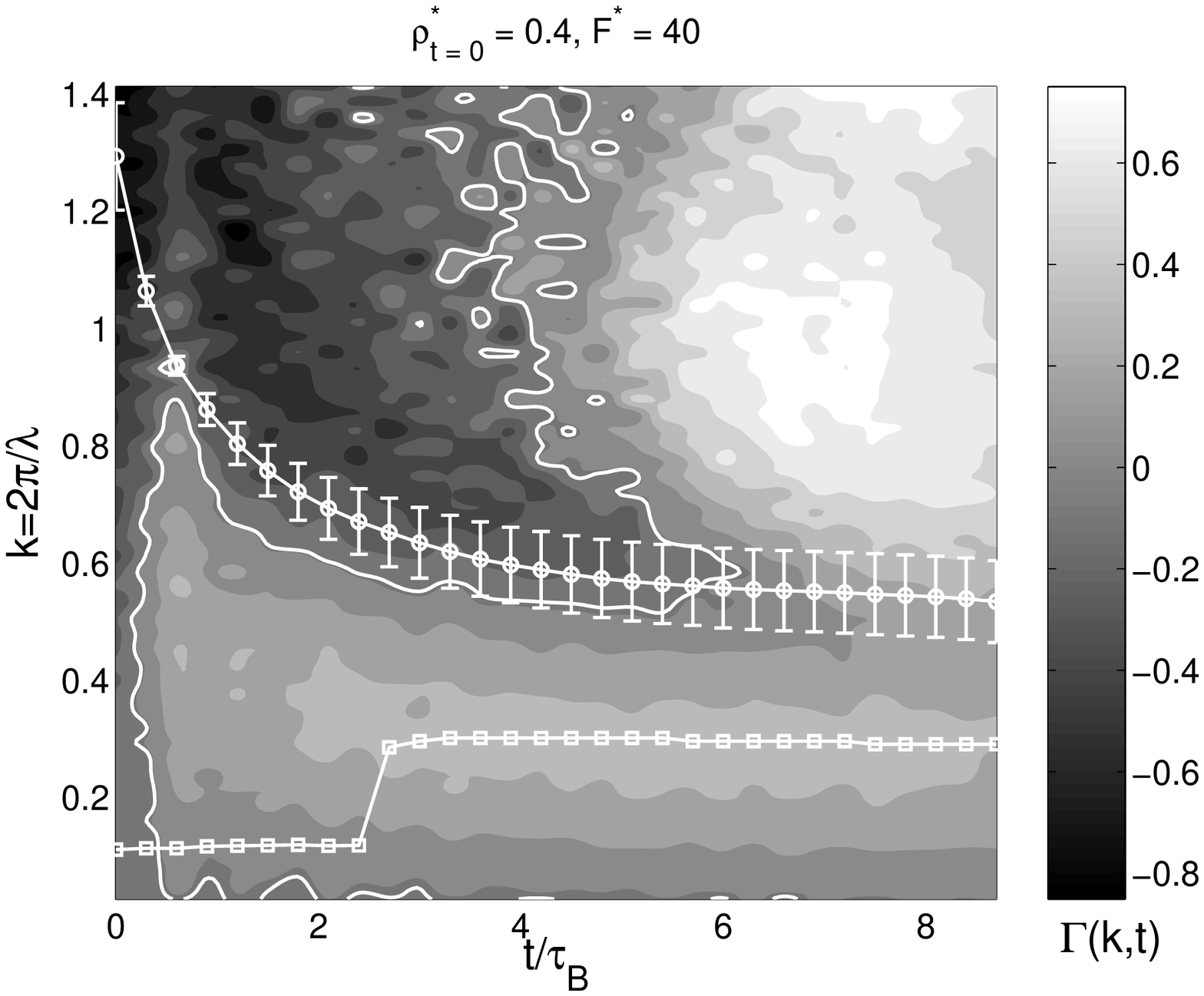,width=14cm}
\label{Figure 4d} 
\end{figure}
\end{center}

\newpage
\begin{center}
\begin{figure}
\epsfig{figure=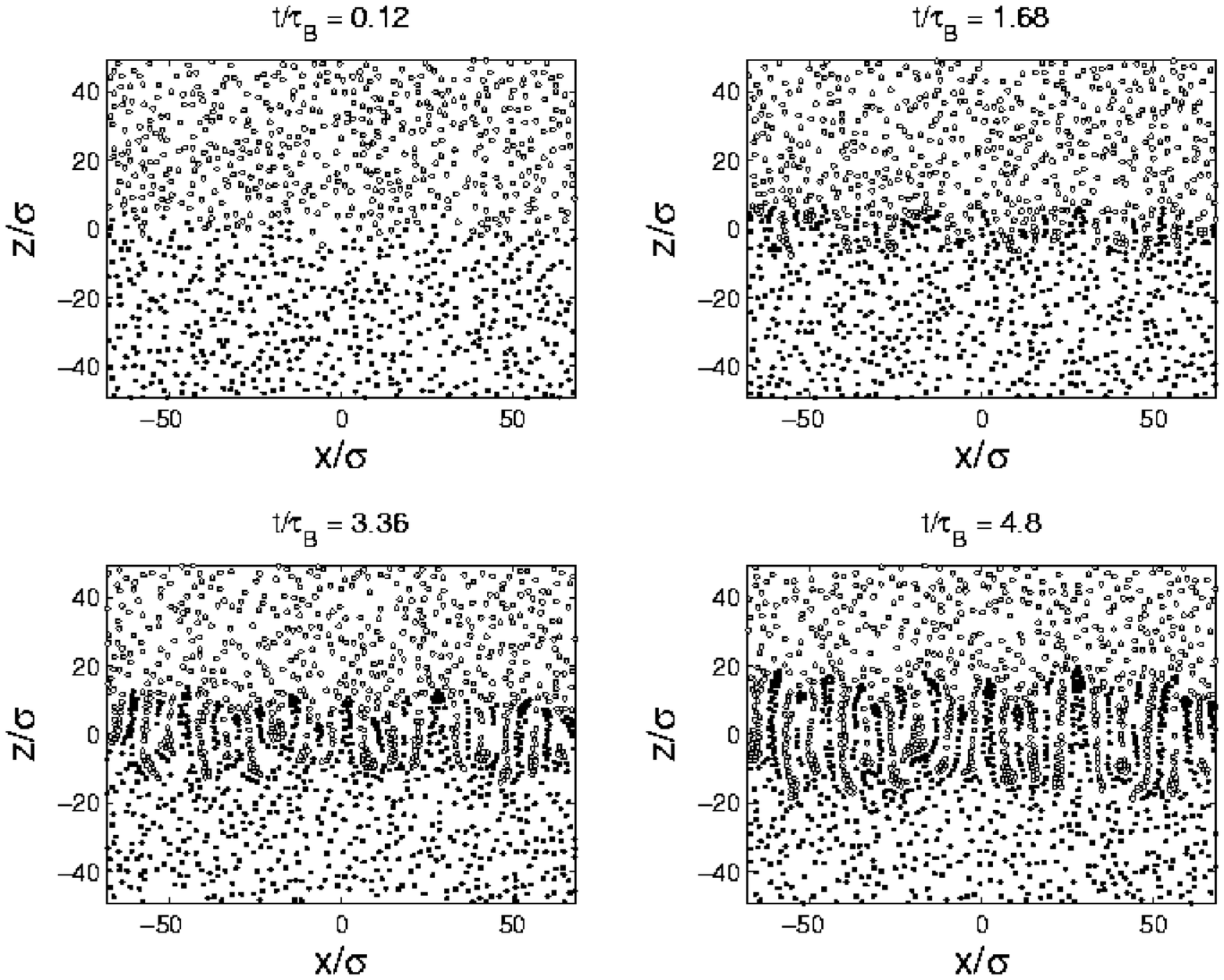,width=14cm}
\label{Figure 6} 
\end{figure}
\end{center}

\end{document}